\documentstyle[epsf,aps,prb,multicol]{revtex}
\begin{document}
\newcommand{\Pvec}{{\rm\bf P}}
\newcommand{\Evec}{{\rm\bf E}}
\newcommand{\eps}{\epsilon}  
\newcommand{\veps}{\varepsilon}  
\newcommand{\De}{$\Delta$}
\newcommand{\de}{$\delta$}
\newcommand{\mc}{\multicolumn}
\newcommand{\be}{\begin{eqnarray}}
\newcommand{\ee}{\end{eqnarray}}
\newcommand{\einf}{\varepsilon^\infty}
\newcommand{\ez}{\varepsilon^0}  

\draft \title{Band offsets and stability of
BeTe/ZnSe (100) heterojunctions}  
\author{Fabio Bernardini,$^{a}$ Maria Peressi,$^{b}$ and Vincenzo
 Fiorentini$^{a}$} 
\address{
$(a)$ INFM and Dipartimento di  Fisica, Universit\`a di Cagliari, Cittadella
 Universitaria, I-09042 Monserrato (CA), Italy\\
$(b)$ INFM and Dipartimento di  Fisica Teorica, Universit\`a di Trieste,
Strada Costiera 11, I-34014 Trieste, Italy}
\date{\today} 
\maketitle

\begin{abstract}
We present ab-initio studies  of band offsets,
formation  energy, and stability of  (100) heterojunctions between
(Zn,Be)(Se,Te) zincblende compounds,
and in particular of the lattice-matched BeTe/ZnSe interface.
Equal band offsets are found at Be/Se and Zn/Te abrupt interfaces, 
as well as at mixed
interfaces, in agreement with the established 
understanding of band offsets at isovalent
heterojunctions. Thermodynamical arguments suggest that  
islands of non-nominal composition  
may form at the interface, 
causing offset variations over $\sim$ 0.8 eV depending 
on growth conditions. Our findings reconcile 
recent experiments on 
BeTe/ZnSe with the accepted theoretical description. 
\end{abstract}


\begin{multicols}{2}
On the basis of experiments as well as theory,
it is commonly accepted that at isovalent semiconductor interfaces
the band offset is almost independent of the local atomic arrangement, 
except in the presence of heterovalent interlayers or 
 antisites.\cite{AFVdW,review} 
This result was originally established for the common-ion systems, 
and later generalized to the whole class of isovalent heterojunctions. 
Investigations on 
no-common-ion systems such as InP/GaInAs and InAs/GaSb  
confirmed that the band offset is independent on the atomic-scale interface 
arrangement,
 despite the different   interface composition and local strain.
\cite{Hybertsen,Peressi-inpgainas,Montanari} Remarkably,
these findings have found a rationale within the linear response theory (LRT)
of band offsets,\cite{review,LRT} which also accounts for 
the composition-dependent local strain effects.\cite{LRT-inasgasb}

Only at heterovalent junctions does the band offset depend  
crucially on interface morphology, due to the different
chemical valence of the atomic constituents, and it is fully explained within 
LRT.
So far, the maximum variation experimentally detected amounts
to 0.6 eV, and it  was observed at  ZnSe/GaAs (100).\cite{Nicolini} 

Controversial findings  have been reported for the isovalent
lattice-matched BeTe/ZnSe (100) interfaces.
In a first paper,\cite{Waag} a valence band offset (VBO) of 0.9 eV was deduced 
from the luminescence spectra of BeTe/ZnSe quantum wells. 
In a subsequent work,\cite{Nagel}
thin BeTe films grown on ZnSe (100) 
were investigated by
XPS: unexpectedly,
two widely different values, 
0.46 eV and 1.26 eV, were measured in different growth conditions,
and interpreted as due to Se- and Zn-terminated substrates.

Should this interpretation be confirmed, this result would be  
a) the first case of morphology-dependent band offset 
at isovalent interfaces; b) the largest VBO variation (0.8 eV)
ever observed at semiconductor heterojunctions;
c) a clear violation of the LRT of offsets.  

To help the interpretation of the
experimental  data, here 
we investigate  the band alignment and the thermodynamical 
stability of BeTe/ZnSe(100), 
and other related  junctions  among (Zn,Be)(Se,Te) compounds,
using 
first-principles 
density-functional-theory
calculations. 
These have proven to be a highly reliable 
tool in predicting offsets at semiconductor-semiconductor 
interfaces. 
  
Pseudopotential plane-waves calculations are performed  
using the VASP code,\cite{vasp}
with ultrasoft pseudopotentials \cite{uspp} for Be and Zn (including
 Zn 3$d$ states in the valence). The generalized gradient approximation 
(GGA-PW91 \cite{pw91})  to the exchange-correlation functional is used.
 All relevant properties of the binary (Zn,Be)(Se,Te) 
compounds are converged at a cutoff of 23
Ry. As shown in Table~\ref{tab.alat}, the experimental 
lattice parameters are reproduced  within 1\%. 
The abrupt interfaces were modeled in periodic 
boundary conditions by 24-atom slab supercells.
For mixed or reconstructed interfaces, 48-atom supercells 
with a total of 24 atomic layers  and 2 atoms per layer were used. 
Brillouin zone integration was performed on a 6$\times$6$\times$2  
Monkhorst-Pack mesh.
The in-plane (substrate) lattice parameter in the supercell 
calculation  was chosen to be a$_{\rm sub}$=5.697 $\rm \AA$, i.e.
 the average of the theoretical bulk lattice parameters of BeTe and
ZnSe. 

For each interface, the supercell structure was fully optimized,
as it is mandatory  to obtain realistic results, since Zn-Te (Be-Se) bond 
lenghts  differ by  about $\pm$ 10\%
from those of bulk BeTe and ZnSe. 
We relax ionic positions and cell parameters until forces below 
0.05 eV/$\rm \AA$ and stress
along the (100) direction lower than 0.5 Kbar are obtained. 
The VBO is then
computed following the approach described in Ref.\onlinecite{review}.
The comparison of the VBO calculated for the ideal unrelaxed 
and for the optimized structures confirm the importance of structural
optimization, whose effects on VBO amount to about 0.4 eV.

Two kinds of abrupt interfaces are possible at BeTe/ZnSe(100)
heterojunctions: the 
Zn/Te interface, characterized by the  sequence of atomic planes
{\dots -Be-Te-Be-{\bf Te}-{\bf Zn}-Se-Zn-Se- \dots}, and the 
complementary Be/Se interface, with the  stacking sequence
{\dots -Te-Be-Te-{\bf Be}-{\bf Se}-Zn-Se-Zn- \dots}~. 
In supercells with periodic boundary conditions, interfaces are always
present in pairs, and they may be chosen (for the present orientation)
to be  different or  identical, depending on the atomic filling of 
the supercell.   We consider supercells both with
identical, symmetry-equivalent interfaces (the first two columns of 
Table~\ref{tab.vbo}) and asymmetric 
supercells with different 
interfaces (the third column of the Table, marked "asym"). The comparison
of the results allows to reduce the numerical uncertainty in the estimate
of the VBO. 
Supercells with symmetry-equivalent interfaces exhibit a non-ideal
$c/a$  ratio  due to local strain
in the interface regions, whereas in supercells with  different and 
complementary interfaces the local positive and negative 
strains nearly compensate, and $c/a$ is close to the ideal unstrained value. 
The average VBO for the abrupt relaxed (100) interfaces
is 0.52 eV, and  all values 
fall within 
a range of  $\sim$ 30 meV, which is of the same order of magnitude
of the numerical uncertainty of the calculations ($\sim$ 10 meV),
in analogy to the results for other no-common ion heterojunctions.
\cite{Hybertsen,Peressi-inpgainas,Montanari}
The equivalence of the two interfaces occurs also for
the ideal unrelaxed  cases.  
Therefore LRT is valid also for this system, despite 
the large chemical differences between the constituting compounds
(e.g., BeTe and ZnSe bulks have a very different ionicity, 0.34
and 0.59 on the Garcia-Cohen scale\cite{GarciaCohen}).
The validity of LRT 
is also confirmed by the fact that,
for the unrelaxed case,   
the VBO for the abrupt (110)
interface is close to the average of the two different (100)
terminations. 
The VBO's
between differently oriented, relaxed interfaces 
shows larger differences;
but this is not in contrast  
with the LRT picture, because local interface strains depend
not only on composition but also on orientation.
From all the above results, we definitely rule out the possibility that 
the difference of about
800 meV between the VBO's measured at BeTe/ZnSe (100) interfaces
can  be simply ascribed to chemically
different abrupt interfaces.\cite{Nagel}

As a further  check,  we studied some prototypical (100)
non-abrupt interfaces, restricting to  either anion- or
cation-intermixed
cases (antisites are generally energetically unfavorable
in II-VI compounds).  In particular we consider {\it c}(2$\times$2) 
reconstructed interfaces with one mixed layer of either 
Be and Zn atoms (fourth column in Table~\ref{tab.vbo}) or Se and Te atoms 
(fifth column in the Table).
Again, the 
VBO is independent of the interface local atomic 
arrangement within  10 meV.
Therefore, we conclude that the VBO at
BeTe/ZnSe (100) heterojunction  does not depend on the interface local
atomic arrangement, thus confirming previous evidence for  isovalent
interfaces, and the general predictions of LRT.  

The DFT-GGA VBO values are not directly comparable with experimental data, 
since spin-orbit coupling and self-energy effects on bulk bands are not taken
into account in this type of electronic structure calculations.
The spin orbit splitting is 0.96 eV for BeTe and 0.40 eV for ZnSe; including 
a posteriori the ensuing correction to the VBO reported in Table~\ref{tab.vbo}, 
we obtain an estimate of about 0.7 eV.
Many-body corrections to valence band top edges, 
still excluded from this estimate, are not 
available for these compounds to our knowledge.
Typical values for these corrections are of order 0.1-0.2 eV,
\cite{zhu-cohen} 
so that a final theoretical estimate could be close
to the experimental value of 0.9 eV reported in Ref.~\onlinecite{Waag}
and, incidentally, to the average of the two values of Ref.~\onlinecite{Nagel}.
However, we stress that the corrections to the DFT-GGA VBO values
are bulk quantities, and thus they affect the {\it absolute} 
value of the VBO, but not at all the {\it relative} comparison  among
the values for different cases considered here. 
Therefore the main result of our
calculations, i.e. the independence of the VBO on interface composition,
is fully valid. 

\begin{figure}[h]
\epsfclipon
\epsfxsize=8cm
\centerline{\epsffile{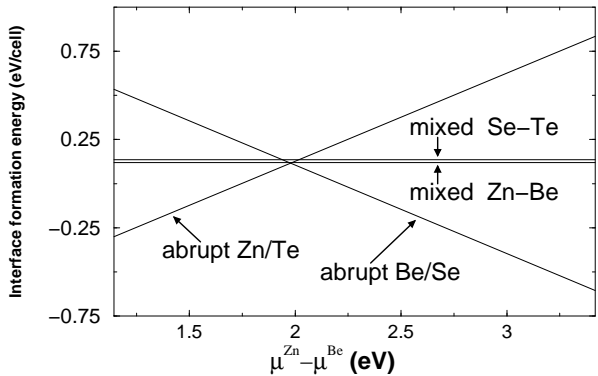}}
\caption{Formation energy (in eV/(1$\times$1) cell) of the abrupt and
c(2$\times$2)-reconstructed 
BeTe/ZnSe (100) interfaces as function of the difference of 
the Zn and Be  chemical potentials.}
\label{fig.abrupt}
\end{figure}

According to our calculations, 
the VBO of 0.9 eV reported in Ref.~\onlinecite{Waag}
could correspond to several possible interface compositions,
including  either  Zn/Te or Be/Se abrupt terminations or mixed interfaces.
However,  some suggestions about the actual structure of the interface
comes from a thermodynamic investigation of interface 
stability.
We find that in thermodynamic equilibrium 
{\it  abrupt interfaces of either kind are favored over the intermixed ones}.
We define the interface formation energy per unit of sectional area
in the most general case  as
\begin{eqnarray}
2\, E^{\rm intf}_{\rm form} & =&
E^{\rm intf}_{\rm tot} - N_{\rm Be} \, \mu^{\rm Be} 
-  N_{\rm Zn} \, \mu^{\rm Zn} 
-  N_{\rm Te} \, \mu^{\rm Te} 
-  N_{\rm Se} \, \mu^{\rm Se} 
\nonumber
\end{eqnarray}
where $E_{\rm tot}^{\rm intf}$ is the total energy of the supercell describing
the interface, and the $\mu$'s and $N$'s are the chemical potentials and 
number of atoms of the various elements involved.
At equilibrium the 
chemical
potentials of the elements and total energies of the condensed phases 
are related by
\begin{equation}
  \rm   \mu^{\rm BeTe} = \mu^{\rm Be} + \mu^{\rm Te}; \ \ \ \ \
  \rm   \mu^{\rm ZnSe} = \mu^{\rm Zn} + \mu^{\rm Se}.
\label{eq.bulks}
\end{equation}
The formation energy of abrupt interfaces is 
easily seen to be a function of the 
difference between Zn (or Se) 
and Be (or Te) chemical potentials. 
Indeed, using Eqs.~\ref{eq.bulks},   
the formation energy for abrupt Zn/Te and Be/Se interfaces
reads
\be
\nonumber
2\, E_{\rm form}^{\rm Zn/Te}& =& E_{\rm tot}^{\rm Zn/Te}
                             -N_{\rm Te}\, \mu^{\rm BeTe}
                             -N_{\rm Se}\, \mu^{\rm ZnSe}
                             - (\mu^{\rm Zn} - \mu^{\rm Be}),\\
2\, E_{\rm form}^{\rm Be/Se}& =& E_{\rm tot}^{\rm Be/Se}
                             -N_{\rm Te}\, \mu^{\rm BeTe}
                             -N_{\rm Se}\, \mu^{\rm ZnSe}
                             + (\mu^{\rm Zn} - \mu^{\rm Be}),
\nonumber
\ee
respectively.
The range of variation of $\mu^{\rm Zn} - \mu^{\rm Be}$ is 
\be
\nonumber
  \rm   
&\mu&^{\rm Zn} - \mu^{\rm Be} \leq
\mu^{\rm Zn-bulk} -\mu^{\rm Be-bulk} - \Delta H^{\rm BeTe}, \\ 
                 &\mu&^{\rm Zn} - \mu^{\rm Be} \geq 
        \mu^{\rm Zn-bulk} - \mu^{\rm Be-bulk} + \Delta H^{\rm ZnSe}, 
\nonumber
\ee 
where $\rm \Delta H^{\rm X}$ is the formation  entalpy for
compound X.
Mixed-interface supercells are instead stoichiometric
($N_{\rm Se}$=$N_{\rm Zn}$=$N_{\rm ZnSe}$, and
$N_{\rm Te}$=$N_{\rm Be}$=$N_{\rm BeTe}$),
therefore the
formation energy is independent of the
chemical potentials.
The previous expression becomes
\be
\nonumber
2\, E_{\rm form}^{\rm mixed} = E_{\rm tot}^{\rm mixed} - N_{\rm BeTe}
\, \mu^{\rm BeTe} -  N_{\rm ZnSe} \, \mu^{\rm ZnSe}
\ee
where now $N$'s and $\mu$'s are referred to the bulk formula unit. 
The results, summarized in Figure \ref{fig.abrupt}, 
show that the Zn/Te abrupt interface  is
favored  in high $(\mu^{\rm Zn} - \mu^{\rm Be})$ conditions, and conversely the 
Be/Se abrupt interface  is
favored  in low $(\mu^{\rm Zn} - \mu^{\rm Be})$ conditions. 
Most interestingly
we  find that, unlike the case of heterovalent junctions,
the present isovalent abrupt  interfaces are always favored over
the mixed ones for the whole range of admissible chemical potentials.
This  behavior was already predicted  for the III-V isovalent GaInP/GaAs 
interface,\cite{fabio} so we suggest that this preference for abrupt
interfaces may be generally valid for  {\it any} isovalent
heterojunction. 

Thermodynamics further gives key 
indications (at least as far as equilibrium energetics is concerned)
on the possible origin of different offsets 
measured in real samples in particular growth conditions.
Islands of {\it a priori} unexpected composition,
 such as BeSe or ZnTe, may form during the deposition of BeTe on
 ZnSe: specifically one expects  BeSe islands
 in Be-rich and Se-rich growth conditions,
and ZnTe islands in  Zn-rich and Te-rich conditions.
In terms of band offsets, the idea  is 
 that these ``hetero-islands'' may in fact be the 
 material effectively interfaced to ZnSe, and therefore largely determine the 
observed band offset. 

\begin{figure}[h]
\epsfclipon
\epsfxsize=8cm
\centerline{\epsffile{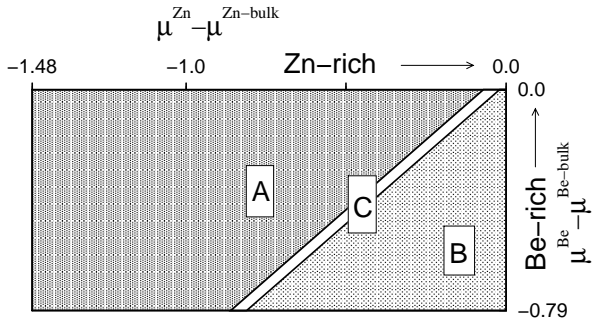}}
\caption{Phase diagram for the (Be,Se,Zn,Te) common-ion interfaces.
The stable interfaces are  BeSe/ZnSe in region A, ZnTe/ZnSe
 in region B, BeSe/ZnTe in region C.}
\label{fig.phase}
\end{figure}

The idea  of islands formation is suggested by previous experience 
with dopant incorporation in semiconductors. 
It was shown theoretically~\cite{Van.de.Walle} 
that rising the
 chemical potential of the Li acceptor
up to its  bulk value, Li incorporation in ZnSe is 
preempted by the formation of a Li$_2$Se surface phase. Indeed,  
 heavy Li doping of  ZnSe layers in MBE growth~\cite{Zhu} results 
in the formation 
of Li$_2$Se islands on the ZnSe surface.
In the present case
 the scenario  is slightly more complex,
 as four
chemical potentials are involved.
We choose as reference the cation
chemical potentials, both for convenience and because  the
cations are the mobile species;
the phase diagram of the four-component interface system will
thus be drawn in the \{$\mu^{\rm Zn}$, $\mu^{\rm Be}$\} plane. 
The reactions leading to the  formation of an epitaxial compound on
 ZnSe at the expenses of BeTe are as follows: for ZnTe on ZnSe,
\be
    \rm Zn + BeTe  \Rightarrow Be + ZnTe,  
\label{uno}
\ee
and  for BeSe on ZnSe
\be
    \rm Be + ZnSe  \Rightarrow Zn + BeSe.
\label{due}
\ee
These reactions will occur exothermically if the
  reaction energy $\rm \Delta E$ is negative; the latter energy 
is given for reactions \ref{uno} and \ref{due}  
by
\be
\nonumber
 \rm   \Delta E^{\rm ZnTe} &=& 
               \mu_{\rm s}^{\rm ZnTe} - \mu^{\rm Be} - \mu^{\rm BeTe}
+  \mu^{\rm Zn},    \\
  \rm  \Delta E^{\rm BeSe} &=& 
                \mu^{\rm BeSe}_{\rm s} - \mu^{\rm Zn} - \mu^{\rm ZnSe} + 
\mu^{\rm Be},
\nonumber
\ee
respectively.
In these relations, $\mu^{\rm XY}_{\rm s}$ is the total energy of
 bulk XY in the {\it pseudomorphically strained} geometry on ZnSe,
 as it results from the optimized XY/ZnSe interface supercell.
Using these equations and the calculated values of the chemical potentials and
compounds formation energies,
we determine the regions in the \{$\mu^{\rm Zn}$,$\mu^{\rm
Be}$\}
plane where  BeTe and ZnSe are unstable
with  respect to  trasformation into ZnTe and BeSe. 
The phase diagram is represented in Fig~\ref{fig.phase}. 
In region A,
$\rm \Delta E^{\rm BeSe}$ is negative
 and
$\rm \Delta E^{\rm ZnTe}$ is positive: therefore the  formation of 
epitaxial BeSe through reaction \ref{due} is energetically favored.
In region B, $\rm \Delta E^{BeSe}$ is positive and 
$\rm \Delta E^{\rm ZnTe}$  
negative, hence epitaxial ZnTe is energetically favored over BeTe. 
In region C, both
the $\Delta {\rm E}$'s are negative, 
hence both BeTe and ZnSe are unstable respect to 
decomposition into BeSe and ZnTe. 
According to this  picture, at thermodynamical equilibrium  
BeTe/ZnSe interfaces {\it are never stable} 
 and the following interfaces may locally form instead:
referring to Fig. \ref{fig.phase}, 
BeSe/ZnSe in region A,
ZnTe/ZnSe in region B, 
and ZnTe/BeSe in the (very small) region C.
Our present result indicates that   interfaces established in real BeTe/ZnSe
samples 
might be locally closer to  ZnTe/ZnSe in Zn-rich conditions and BeSe/ZnSe
in Be-rich conditions, than to the nominal BeTe/ZnSe composition. 
A direct consequence of this result which should be
observable in experiment is the preferential formation
of BeSe or ZnTe islands on ZnSe during the early
stages of growth of a nominally BeTe-ZnSe interface.
Our analysis 
does not include growth kinetics effects,
which may cause the
(unstable)
nominally-BeTe/ZnSe interface to actually form 
for chemical potentials in region C of Fig. \ref{fig.phase},
where the thermodynamic driving force towards equilibrium
(i.e. instability of BeTe/ZnSe)  is smallest.

\begin{figure}[h]
\epsfclipon
\epsfxsize=8cm
\centerline{\epsffile{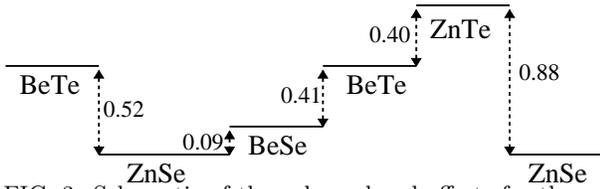}}
\caption{Schematic of the valence band offsets for the various
interfaces
investigated.}
\label{fig.vbo1}
\end{figure}

We now discuss the  key piece of information we are looking for,
namely  the VBO values
 for the various possible interfaces. In calculating them, we use the
same in-plane lattice parameter a$_{\rm sub}$  as in 
 all previous calculations (the substrate is unchanged),
and carefully account for bulk and interfacial strain effects.
As in the BeTe/ZnSe case, the calculated VBO values
are affected by a substantial {\it absolute} uncertainty
due to many-body and spin-orbit splitting effects,
here combined with splittings coming from epitaxial strain.
However, the {\it relative} uncertainty
in comparing the values for the different systems 
is much smaller, due to a partial cancellation of systematic 
corrections to the  bulk band edges.
The results are 
 depicted schematically (also including the no-common-ion 
lattice-matched interface)
in  Fig.~\ref{fig.vbo1}. Two points are relevant in the Figure:
{\it (a)} the 
transitivity  rule \cite{review} holds within the numerical
uncertainty of the calculations,   confirming once again the validity
of LRT for these systems;
{\it (b)} the values of the VBO for the different systems differ at most
by  about 0.8 eV,
the minimum value corresponding to the BeSe/ZnSe interface (Be-rich
conditions) and the
maximum to ZnTe/ZnSe (Zn-rich conditions).

It is interesting to note that the maximum calculated VBO difference of  0.8 eV
is the same as the one  measured in the two different samples in
 Ref.~\onlinecite{Nagel}; in addition, in that experiment the maximum value 
was observed in Zn-rich conditions, and the  minimum value
in Se-rich conditions, in agreement with
 our findings. This matching 
suggests a possible correspondence
between the  lower (higher) experimentally measured VBO \cite{Nagel} and 
the formation of a BeSe/ZnSe (ZnTe/ZnSe) interface, 
although
the absolute values of the  calculated offsets are about
0.3 eV lower than the measured ones, because of the discussed unaccuracy
of the theoretical estimate. 

In conclusion,
we presented band offset calculations for a series of zincblende 
(100)  interfaces between various (Zn,Be)(Se,Te)
II-VI compounds. We also  set up a
 thermodynamical phase diagram bearing on the stability of the 
various possible interfaces. Based on our results, we discussed
recent experiments
on BeTe/ZnSe interfaces, which showed a marked offset variation
with growth conditions. 
Our conclusions are  that: $(i)$ the attribution of the two widely different 
measured VBO values to  abrupt Se-terminated and Zn-terminated
interfaces of the nominal BeTe/ZnSe heterojunction, as proposed in 
Ref.~\onlinecite{Nagel}, is incorrect, as well as any other
attribution to mixed (reconstructed) interfaces, which
have a composition-independent VBO;
$(ii)$ conversely, 
strained interfaces between other (Zn,Be)(Se,Te) compounds shows
a VBO which may differ up to 0.8 eV in the case of
BeSe/ZnSe and ZnTe/ZnSe; $(iii)$ 
thermodynamics indicates that 
such interfaces may actually locally
form in the deposition of BeSe on ZnTe and viceversa;
$(iv)$ interfaces between (Zn,Be)(Se,Te) compounds follow closely the linear 
reponse theory predictions just as III-V--based systems.
Although  the problem require further investigation for a definite
explanation, the comparison of experimental and 
our theoretical findings could  suggest that
observed interfaces may locally be {\it not}
the nominal BeTe/ZnSe, but rather interfaces such as BeSe/ZnSe or
ZnTe/ZnSe depending on the chosen growth conditions. 

We acknowledge support from 
Istituto Nazionale per la Fisica della Materia under 
 the ``Iniziativa Trasversale di Calcolo Parallelo''.
We are in debt to N. Binggeli and A. Franciosi for useful discussions.

\end{multicols}

\narrowtext
 
\narrowtext

\narrowtext

\narrowtext
\begin{table}[h]
\caption{Theoretical and experimental \protect\cite{Madelung}
lattice parameters  (in \AA) of the zincblende
compounds considered in this work.}
\begin{tabular}{lcccc}
Compound     & BeTe    & BeSe         & ZnSe         & ZnTe   \\ 
\hline
Exp.         & 5.626   & 5.139        & 5.668        & 6.104  \\
GGA-PW91     & 5.661   & 5.174        & 5.734        & 6.182    
\end{tabular}
\label{tab.alat}
\end{table}

\narrowtext
\begin{table}[h]
\caption{Valence-band offset (in eV) at different BeTe/ZnSe interfaces.
The  valence band top edge is higher in  BeTe than in ZnSe.}
\begin{tabular}{lcccccc}
            & \mc{3}{|c|}{(100) abrupt}  
            & \mc{2}{c|}{\parbox{22mm}{\centering (100)~c(2$\times$2)}}
                                        & \mc{1}{c}{(110)} \\
\multispan{1}{\hskip\tabcolsep\smash{\parbox{14mm}{\def\baselinestretch{.85}
\normalsize       \centering
Structure}}\hskip\tabcolsep}
&\multispan{5}{}\\  
          & \mc{1}{|c}{Zn/Te} & \mc{1}{c}{Be/Se} & \mc{1}{c|}{asym}  
          & \mc{1}{c}{~Zn-Be} 
          & \mc{1}{c|}{Se-Te} & 
\multispan{1}{\hskip\tabcolsep\smash{\parbox{12mm}{\def\baselinestretch{.85}
\normalsize    \centering 
abrupt }} \hskip\tabcolsep}\\ 
\hline
ideal     & 0.16  & 0.13  & 0.15    &  0.12       &  0.12         & 0.14 \\ 
relaxed   & 0.54  & 0.51  & 0.51    &  0.53       &  0.52         & 0.59
\end{tabular}
\label{tab.vbo}
\end{table}

\end{document}